\documentclass[preprint]{aastex}
\usepackage{subfigure}
\pdfoutput=1

\slugcomment{}

\shorttitle{Astrometry of Cassini}
\shortauthors{Jones et al.}

\begin{document}

\title{Astrometry of Cassini with the VLBA to \\
    Improve the Saturn Ephemeris}

\author{Dayton L.~Jones}
\affil{Jet Propulsion Laboratory, California Institute of Technology, 
Pasadena, CA 91109}
\email{dayton.jones@jpl.nasa.gov}

\author{William M.~Folkner}
\affil{Jet Propulsion Laboratory, California Institute of Technology, 
Pasadena, CA 91109}

\author{Robert A.~Jacobson}
\affil{Jet Propulsion Laboratory, California Institute of Technology, 
Pasadena,CA 91109}

\author{Christopher S~Jacobs}
\affil{Jet Propulsion Laboratory, California Institute of Technology, 
Pasadena,CA 91109}

\author{Vivek Dhawan}
\affil{National Radio Astronomy Observatory, Socorro, NM 87801}

\author{Jon Romney}
\affil{National Radio Astronomy Observatory, Socorro, NM 87801}

\and

\author{Ed Fomalont}
\affil{National Radio Astronomy Observatory, Charlottesville, VA
22903}

\begin{abstract}
Planetary ephemerides have been developed and improved over centuries.  They are a fundamental tool for understanding solar system dynamics, and essential for planetary and small body mass determinations, occultation predictions, high-precision tests of general relativity, pulsar timing, and interplanetary spacecraft navigation.  This paper presents recent results from a continuing program of high-precision astrometric very-long-baseline interferometry (VLBI) observations of the Cassini spacecraft orbiting Saturn, using the Very Long Baseline Array (VLBA).  We have previously shown that VLBA measurements can be combined with spacecraft orbit determinations from Doppler and range tracking and VLBI links to the inertial extragalactic reference frame (ICRF) to provide the most accurate barycentric positions currently available for Saturn.  Here we report an additional five years of VLBA observations along with improved phase reference source positions, resulting in an improvement in residuals with respect to the Jet Propulsion Laboratory's dynamical ephemeris.  
\end{abstract}

\keywords{techniques: interferometric --- astrometry --- planets and satellites: individual (Saturn)}

\section{Introduction}

For the past decade, observations of Cassini have been carried out by the Jet Propulsion Laboratory (JPL) and the National Radio Astronomy Observatory (NRAO) with the Very Long Baseline Array (VLBA)\footnote{The VLBA is a facility of the National Radio Astronomy Observatory (NRAO), which is operated by Associated Universities, Inc., under a cooperative agreement with the National Science Foundation.} (\citet{Jones11}; \citet{Jones12}).  These observations are planned to continue until the end of the Cassini mission in 2017. We will then have high accuracy measurements over the largest possible fraction of SaturnÕ's orbital period.  The error in determining the plane of SaturnÕs orbit (latitude) decreases rapidly as the total time span of observations increases, and a longer total time span for the astrometric observation allows the ephemeris improvements to be extended much farther into the future.  Accurate ephemerides are one of the basic tools of astronomy, whose accuracy requires regular observational support to maintain and improve.  As an example, NANOGrav and other pulsar timing arrays (e.g., Perrodin, et al. 2013) will directly benefit from future ephemeris improvements.  

The orbits of the inner planets are accurately tied together with current data, but the outer planets are not as well tied to the inner planets or each other.  The Cassini mission provides our first opportunity to incorporate high-accuracy data from a spacecraft orbiting an outer planet for an extended period of time\footnote{The Galileo spacecraft orbited Jupiter for several years, but the failure of its high gain antenna to deploy severely limited the accuracy of VLBI tracking.}.  
Our goal is to improve the accuracy of barycenter position measurements of Saturn in the inertial International Celestial Reference Frame (ICRF and ICRF2, Fey et al. 2009; Ma et al. 2009) through phase-referenced very-long-baseline interferometry (VLBI) observations of Cassini at 8.4 GHz combined with orbit determinations from Doppler and range tracking.  (See \citet{Asmar05} for a summary of spacecraft Doppler tracking.)  The Cassini orbit can be determined to about 2 km at apoapse and 0.1 km at periapse relative to the center of mass of Saturn.  The exact accuracy of Cassini orbit solutions varies from orbit to orbit, as the orbital period changes and shorter orbits provide fewer Doppler measurements per orbit segment.  Previous work has demonstrated a reduction in ephemeris residuals for Saturn of a factor of three by combining observations made with the VLBA and Cassini spacecraft orbit solutions from Deep Space Network (DSN) tracking.     

The stability of the ICRF2 (rotational accuracy) is approximately 0.01 milli-arcsec (mas; 1 mas = 5 nrad), although individual source may have position errors of 0.05-0.15 mas or more (\citet{Fey04}; \citet{Porcas09}; \citet{Ma09}; \citet{Fey09}) due to changes in source morphology.  The high resolution imaging capability of the VLBA allows us to detect source changes.  Prior phase referenced VLBI astrometry results (e.g., \citet{Lestrade99}) suggest that a precision of 0.1 mas with respect to nearby ICRF2 sources can be achieved for Cassini.  Thus, the combined error of our absolute Cassini positions with respect to the ICRF2 is expected to be about 0.2 mas, similar to VLBI spacecraft tracking errors calculated by \citet{Lanyi07}, \citet{Border08}, and \citet{Curkendall13}.

\section{New Observations}

We have observed Cassini with the VLBA during twelve epochs between 2008 and 2014 under experiment codes BJ061, BJ067, BJ079, and BJ082.  Results from the first eight epochs were reported in \citet{Jones11}.  Table \ref{tab1} lists the observing epochs up to early 2014, including the VLBA antennas used during each epoch. 
The VLBA consists of ten 25-m diameter radio antennas located in the northern hemisphere from the US Virgin Islands to Hawaii.  It has demonstrated a uniquely good astrometric precision of $<$ 0.01 mas (10 $\mu$as) in particularly favorable circumstances (e.g., \citet{Fomalont03}).   

We used standard phase-referencing techniques (\citet{Counselman72}; \citet{Shapiro79}; \citet{Lestrade90}; \citet{Guirado97};  \citet{Guirado01}; \citet{Fomalont06}) with rapidly alternating scans between Cassini and angularly nearby reference sources (see Table \ref{tab2}).  
Each of our observing epochs was four hours long, including both the alternating short scans on Cassini and phase reference sources and a period of about 40 minutes during which we observed multiple strong sources spread over the sky to allow better fitting for the tropospheric delay at each site (\citet{Lestrade04}; \citet{Mioduszewski04}; \citet{Fomalont05}).  We used an instrument configuration that provided either four separate frequency bands in both right and left hand circular polarizations (RCP and LCP), or eight frequency bands in RCP only.  In both cases we spaced the multiple receiving bands non-uniformly across a wide frequency range of several hundred MHz centered near 8.4 GHz to allow accurate group delay measurements.  During our most recent three epochs we were able to benefit from an upgrade in the VLBA data recording rate to 2 Gb/s from 512 Mb/s.  This effectively doubled the array sensitivity and allowed us to utilize weaker but angularly closer reference sources to improve the cancellation of troposphere delay errors.  All data were processed by the DiFX software correlator (\citet{Deller07}; \citet{Deller11}) at NRAO in Socorro, NM.  

\begin{deluxetable}{ccc}
\tablewidth{0pt}
\tablecaption{Observing Epochs and VLBA Antennas Used \label{tab1}}
\tablehead{
\colhead{Epoch} & \colhead{Obs.~Date} & \colhead{VLBA Antennas (see note)}}    
\startdata    
BJ061A & 2006 Oct 11 & SC, HN, NL, FD, LA, PT, KP, OV, MK \\  
BJ061B & 2007 Mar 1 & SC, HN, NL, FD, LA, PT, KP, OV, BR, MK \\  
BJ061C & 2007 Jun 7 & SC, HN, NL, FD, LA, PT, KP, OV, BR, MK \\
BJ061D & 2008 Jan 12 & SC, HN, NL, LA, PT, KP, OV, BR, MK \\
BJ061E & 2008 Jun 7 & SC, HN, FD, LA, KP, OV, MK \\  
BJ061F & 2008 Aug 1 & SC, HN, NL, FD, LA, PT, KP, OV, BR, MK \\
BJ061G & 2008 Nov 11 & HN, NL, FD, LA, PT, KP, OV, BR, MK \\
BJ061H & 2009 Apr 24 & SC, HN, NL, FD, LA, PT, KP, OV, BR, MK \\
BJ067A & 2009 Jun 24 &SC, HN, FD, LA, PT, KP, OV, BR, MK \\
BJ067B & 2010 Sep 8 & No Cassini signal received \\
BJ067C & 2011 Feb 21 & SC, HN, NL, FD, LA, PT, KP, OV, BR, MK \\
BJ067D & 2012 Feb 5 & SC, HN, NL, FD, LA, PT, KP, OV, BR, MK \\
BJ079A & 2012 Jul 7 & Calibrator position improvement \\
BJ079B & 2013 Mar 31 & SC, HN, NL, FD, LA, PT, KP, OV, BR, MK \\
BJ079C & 2013 Jun 14 & SC, HN, NL, LA, KP, OV, BR, MK \\
BJ079D & 2013 Oct 29 & Observed Mars orbiters only \\
BJ079E & 2014 Jan 6 & SC, HN, FD, LA, PT, KP, OV, BR, MK \\
\enddata
\tablecomments{The VLBA antenna locations are:  SC = St.~Croix, US Virgin Islands; 
HN = Hancock, NH; NL = North Liberty, IA; FD = Fort Davis, TX; LA = Los Alamos, NM; 
PT = Pie Town, NM; KP = Kitt Peak, AZ; OV = Owens Valley, CA; BR = Brewster, WA; 
MK = Mauna Kea, HI.}
\end{deluxetable}
    
\begin{deluxetable}{ccccc}
\tablewidth{0pt}
\tablecaption{Observing Epochs and Phase Reference Sources \label{tab2}}
\tablehead{
\colhead{Epoch} & \colhead{Date} & \colhead{Reference Source} & 
\colhead{Angular Separation} & \colhead{Flux Density} \\
 & & & \colhead{(deg)} & \colhead{(Jy)}}    
\startdata    
BJ061A & 2006 Oct & J0931+1414 & 2.5 & 0.2 \\
BJ061B & 2007 Mar & J0931+1414 & 2.0 & 0.2 \\
BJ061C & 2007 Jun & J0931+1414 & 2.0 & 0.2 \\
BJ061D & 2008 Jan & J1025+1253 & 3.5 & 0.5 \\
BJ061E & 2008 Jun & J1025+1253 & 1 & 0.5 \\
BJ061F & 2008 Aug & J1025+1253 & 3.0 & 0.5 \\
BJ061G & 2008 Nov & J1127+0555 & 0.5 & 0.1 \\
BJ061H & 2009 Apr & J1112+0724 & 0.5 & 0.2 \\
BJ067A & 2009 Jun & J1058+0133 & 6.7 & 2.0 \\
BJ067A & 2009 Jun & J1112+0724 & 0.07 & 0.2 \\
BJ067A & 2009 Jun & J1118+1234 & 5.5 & 0.2 \\
BJ067A & 2009 Jun & J1127+0555 & 4.2 & 0.1 \\
BJ067C & 2011 Feb & J1304-0346 & 0.3 & 0.2 \\
BJ067D & 2012 Feb & J1354-1041 & 1.9 & 0.4 \\
BJ079B & 2013 Mar & J1434-1146 & 0.6 & 0.1 \\
BJ079C & 2013 Jun & J1408-0752 & 3.2 & 0.7 \\
BJ079E & 2014 Jan & J1507-1652 & 2.2 & 0.5  \\
\enddata
\tablecomments{Flux densities listed are average correlated flux densities on the longest VLBA baselines.  Four phase reference sources were used during epoch BJ067A.}
\end{deluxetable}

Data from each epoch were used to produce a phase-referenced image of the spacecraft signal without self-calibration, using the Astronomical Image Processing System (AIPS)\footnote{AIPS is provided and supported by the National Radio Astronomy Observatory.} for data editing, calibration, fringe fitting, and image formation and deconvolution.   
The position of the signal peak was measured and the image shifted to the nominal phase center.  This position shift, plus any residual position error measured from the post-shift baseline phases, was combined with the VLBI geometric model to produce total phase delay data.  The difference in the total delays between the spacecraft and reference source is the observable used by JPL navigation and ephemeris software.

The analysis of data from a phase-referenced VLBI experiment involves the removal of multiple sources of error (e.g., \citet{Lanyi05}).  The following subsections describe the more important of these corrections. 

\subsection{Experiment Scheduling}

Because the apparent motion of Saturn on the sky reverses direction twice every year, it is possible with careful scheduling to use the same phase reference source during multiple epochs.  This reduces the number of separate phase reference sources that need to be tied to the ICRF.  It was not always possible to find a suitable phase reference source within $2^{\circ}$ of Cassini because we were constrained to observe during times when Cassini had its high gain antenna pointing toward Earth and was transmitting.  We used schedules of Cassini tracking passes at the Deep Space Network (DSN) Goldstone complex in California to determine when a signal would be present.  There is also a tradeoff between the angular distance to a particular reference source and its nominal flux density.  The increased VLBA data rate will help in this regard, allowing useful data to be obtained from references source with flux densities well below 100 mJy (our previous flux density cutoff).  In addition to the systematic error cancellation advantages of using angularly closer reference sources, there is evidence that weaker radio sources have morphologies that are more dominated by single compact components \citep{Deller14}. 

\subsection{A Priori Calibration}

The geometric model used during correlation \citep{Romney99} provided the initial delays.  However, the {\it a priori} spacecraft position and proper motions available before each epoch were not sufficiently accurate for correlation, so reconstructed orbit files from JPL were used after each epoch to improve the geometric model.  The model takes into account the difference in general relativity corrections for signals propagating different distances through gravitational fields in the solar system (only part of the solar system gravitational field applies to signals from Cassini).  

Initial amplitude calibration was based on continuously recorded system temperatures and previously determined gain curves for each VLBA antenna.  Corrections were applied for the two-bit signal quantization used at the VLBA antennas, and phase corrections were applied to account for changing parallactic angles and for up-to-date Earth orientation parameters (UT1 and polar motion).  Fixed delay offsets in the electronics were corrected by fringe fitting a strong calibration source and using the resulting delay 
corrections to align phases within each frequency band.  The data were examined for manually to verify that {\it a priori} calibration had been applied correctly.  

\subsection{Ionosphere Delay Calibration}

The frequency differences between our frequency bands were not large enough to accurately calculate dispersive ionosphere delays.  Instead, these delays were calculated from Global Ionosphere Maps (GIM) determined from a large network of GPS receivers at two-hour intervals \citep{Mannucci98}.  For each antenna location the zenith total electron content values from the global maps were linearly interpolated between maps on either side of a particular observing scan.  In addition, a weighted longitude correction was introduced to account for the expected movement of ionospheric features with the Sun during the period between GIM epochs.  

\subsection{Troposphere Delay Calibration}

Troposphere delay calibration has been especially important for the more recent observing epochs, as Saturn has been moving south in declination for the past several years. 
For a northern hemisphere VLBI array that means that observations must be made at lower elevation angles where troposphere effects are enhanced.  During each epoch a sample of approximately 15 strong compact sources covering a wide range of elevation angles at each VLBA antenna were observed in a rapid sequence.  By fitting a linear phase slope to the frequency channels we obtained multi-band delays for each source.  The \citet{Chao74} troposphere delay mapping function was then fit to the multi-band delays to determine the zenith troposphere delay, clock offset, and clock rate for each antenna (see \citet{Sovers98}; \citet{Mioduszewski04}).  When the troposphere delay was properly calculated and removed, the phases were aligned between all frequency bands for each source.  We have not used multiple calibration sources bracketing the position of Cassini to further improve the troposphere delay corrections, as demonstrated by \citet{Fomalont05}, because we wished to keep the beam switching cycle time as short as possible.  Future observations will explore the trade between these observing approaches.  

\subsection{Bandpass Calibration}

The bandpass amplitude and phase response, and any remaining antenna-based residual delays, were corrected by observing a strong calibration source during each epoch.  This is important for our measurements because the spacecraft signal occupies only a small fraction of the bandwidth containing the reference source signal.  Uncalibrated bandpass phase variations would produce a systematic phase offset between the two signals.  After all calibration steps up to this point were applied to the data they were then averaged over single scans and the phases examined to for each source and baseline to verify that they were constant within and between frequency bands. 

\subsection{Phase-Referencing}

\begin{figure}[!h]
\hskip 60pt \includegraphics[angle=0,scale=0.50]{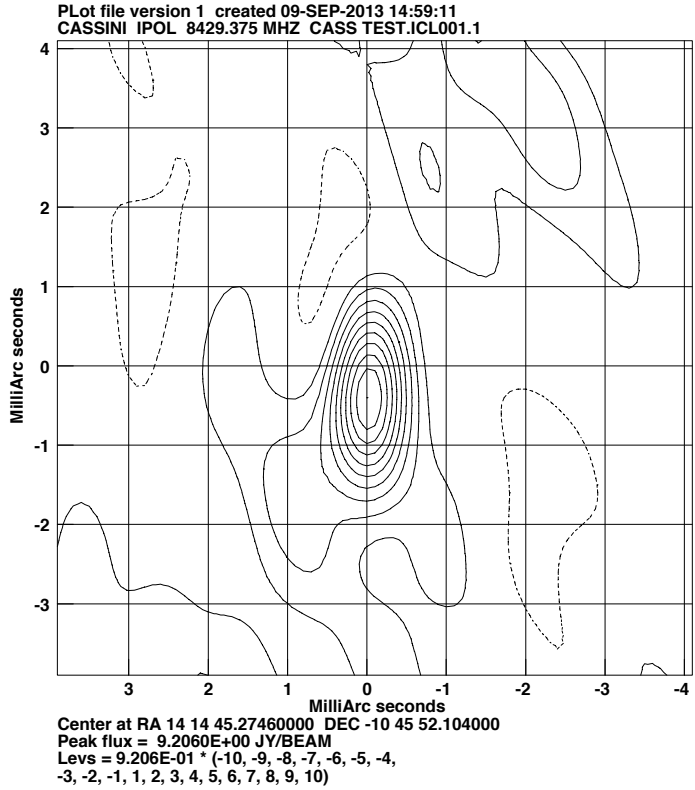}
\caption{Example of a phase referenced image of Cassini showing an offset from the {\it a priori} position based on the phase reference source position and geometric model including the orbital motion of Cassini.  No self-calibration has be applied to the Cassini data.
\label{fig1}}
\end{figure}

We used a point source model for phase self-calibration of the visibilities for the phase reference calibration sources.  The phase corrections from self-calibration were then applied to the Cassini visibilities, removing most of the remaining common errors.  The Cassini data were not self-calibrated.  Images were made of both the phase reference source and Cassini.   The reference source map was examined be check that the point source model had been an adequate approximation.  The peak of the Cassini image was usually visibly offset from the image phase center.  Figure \ref{fig1} shows a typical phase-referenced image of Cassini. 

A two-dimensional quadratic fit was used to measure the position of the Cassini peak, and an appropriate phase slope was applied to the Cassini visibilities to shift the peak to the phase center of the image.  The Cassini baseline phases were examined after the position shift to verify that they were (nearly) flat.  The position shift applied to the Cassini image is the offset in the {\it{a priori}} Cassini position with respect to the {\it{a priori}} position of the phase reference source and the geometric model used.  Consequently, the uncertainly in the ICRF position of Cassini at any epoch can be no better than the uncertainty in the ICRF2 position of the associated reference source, even though the relative position offset between Cassini and the reference source may have a much smaller uncertainty.  Thus, the accuracy of our phase reference source ICRF2 positions is an important issue.  Continuing observations of ICRF2 catalog source positions are constantly improving the accuracies of these positions (e.g, \citet{Fey09}; \citet{Ma09}).  We also use source positions obtained by the DSN (\citet{Border09}; \citet{Jacobs13}), and by GSFC and \citet{Petrov12}.  Ideally we would focus on reference sources that display persistent interstellar scintillation, as these appear to have a more compact structure and higher astrometric positional stability \citep{Schaap13}.  However, the relatively small number of potential phase reference sources precludes using this additional selection filter. 

Table \ref{tab3} shows the positions used for our primary phase reference sources.  Most of these positions are from the ICRF2 or DSN source catalogs, depending on which appeared to have the best accuracy for each source.  Note that some of the sources in Table \ref{tab3} have slightly different positions and smaller errors compared to the positions used by \citet{Jones11}.  Note also that despite the continuing improvements in positional accuracy, four of the ten sources in Table \ref{tab3} still have errors (slightly)  greater than 0.2 mas in declination.  All errors are less than 0.2 mas in right ascension.  The source J1217-0029 does not appear in Table \ref{tab2} because it was used during epoch BJ067B, which occurred during a period when Cassini was not transmitting to Earth.  It is included in Table \ref{tab3} for completeness.
    
\begin{deluxetable}{ccc}
\tablewidth{0pt}
\tablecaption{Primary Phase Reference Sources \label{tab3}}
\tablehead{
\colhead{Source}  & \colhead{RA (J2000)}  &
\colhead{DEC (J2000)}}    
\startdata    
J0931+1414 & 09$^{\rm h}$31$^{\rm m}$05$^{\rm s}$.342427 $\pm0^{\rm s}.000013$ & +14$^{\circ}$14\arcmin16.51851\arcsec  $\pm$0.00023\arcsec \\
J1025+1253 & 10$^{\rm h}$25$^{\rm m}$56$^{\rm s}$.285370 $\pm0^{\rm s}.000004$ & +12$^{\circ}$53\arcmin49.02201\arcsec  $\pm$0.00008\arcsec \\
J1112+0724 & 11$^{\rm h}$12$^{\rm m}$09$^{\rm s}$.558525 $\pm0^{\rm s}.000013$ & +07$^{\circ}$24\arcmin49.11840\arcsec  $\pm$0.00038\arcsec \\
J1127+0555 & 11$^{\rm h}$27$^{\rm m}$36$^{\rm s}$.525539 $\pm0^{\rm s}.000007$ & +05$^{\circ}$55\arcmin32.05913\arcsec  $\pm$0.00012\arcsec \\
J1217-0029 & 12$^{\rm h}$17$^{\rm m}$58$^{\rm s}$.729044 $\pm0^{\rm s}.000008$ & -00$^{\circ}$29\arcmin46.29989\arcsec $\pm$0.00018\arcsec \\
J1304-0346 & 13$^{\rm h}$04$^{\rm m}$43$^{\rm s}$.642222 $\pm0^{\rm s}.000007$ & -03$^{\circ}$46\arcmin02.55176\arcsec $\pm$0.00023\arcsec \\
J1354-1041 & 13$^{\rm h}$54$^{\rm m}$46$^{\rm s}$.518685 $\pm0^{\rm s}.000002$ & -10$^{\circ}$41\arcmin02.65616\arcsec $\pm$0.00003\arcsec \\
J1408-0752 & 14$^{\rm h}$08$^{\rm m}$56$^{\rm s}$.481200 $\pm0^{\rm s}.000001$ & -07$^{\circ}$52\arcmin26.66650\arcsec $\pm$0.00002\arcsec \\
J1434-1146 & 14$^{\rm h}$35$^{\rm m}$21$^{\rm s}$.135833 $\pm0^{\rm s}.000009$ & -11${^\circ}$46\arcmin19.51243\arcsec $\pm$0.00022\arcsec \\
J1507-1652 & 15$^{\rm h}$07$^{\rm m}$04$^{\rm s}$.786962 $\pm0^{\rm s}.000002$ & -16$^{\circ}$52\arcmin30.26701\arcsec $\pm$0.00004\arcsec \\
\enddata
\tablecomments{Positions and errors are from \citet{Fey09}, \citet{Jacobs13}, \citet{Petrov12}, and the GSFC catalog (http://gemini.gsfc.nasa.gov/solutions/2014a/2014a.html).}
\end{deluxetable}

ICRF and DSN source positions are based on group delay measurements, while our Cassini astrometry is based on phase delay measurements.  Source positions measured with group delays are less affected by variations in opacity along the inner regions of radio jets.  Consequently there can be time-variable offsets between group delay and phase delay measurements of the position of the radio centroid of a given reference source.  Investigation by \citet{Porcas09} has found that the offsets between group and phase delay positions are normally less than 0.2 mas at 8.4 GHz.  This is similar to the errors expected from imperfect troposphere delay calibration and individual ICRF2 source position errors, so this effect does not dominate our experimental error budget.  

\subsection{Total Delays}

Total delays for use with JPL navigation and ephemeris software were calculated from AIPS data tables containing the correlator geometric model, the measured residual delays (Cassini image position shift), and any additional small delay corrections from the post-shift Cassini baseline phases.  Calculation of the total delays and the creation of output data files in the format needed by JPL was done with a program written by E.~Fomalont.

\section{Results}

The derived J2000 (ICRF2) positions of the Saturn system barycenter from VLBA observations of Cassini, including detailed Cassini orbit reconstructions, are listed in Table \ref{tab4}.  This table is the main result of our observations.  The orbital solutions for Cassini were produced by integrating the equations of motion as part of a global Saturn ephemeris and gravity field solution (e.g., \citet{Antreasian06}; \citet{Jacobson06}).  These solutions used a large number of recent and historical observations, and include the gravitational effects of solar system objects (including the mutual interactions of Saturnian moons), relativistic perturbations, Saturn oblateness, and non-gravitational effects (spacecraft attitude control, trajectory maneuvers, and solar radiation pressure).  Table \ref{tab4} includes two epochs from VLBA experiment BR103 in 2004, and one epoch from \citet{Fomalont10} in 2009 February.  These experiments were included with the epochs from experiments BJ061, BJ067, BJ079, and BJ082 because they used the same observing technique and instrumentation to determine astrometric positions for Cassini. 

\begin{deluxetable}{cccccc}
\tablewidth{0pt}
\tablecaption{Observed Saturn Barycenter Positions in ICRF 2.0 Reference Frame \label{tab4}}
\tablehead{
\colhead{Date} & \colhead{Time} & \colhead{Observed} & \colhead{Observed} &
\colhead{Error in} & \colhead{Error in} \\
\colhead{ } & \colhead{(UTC)} & \colhead{Right Ascension} & 
\colhead{Declination} & \colhead{R.A. (s)} & \colhead{Dec. (\arcsec)}}
\startdata
2004\ Sep\ 08 & 18:00:00 & 07$^{\rm h}$43$^{\rm m}$57$^{\rm s}$.853971 & +21$^{\circ}$06\arcmin11.47271\arcsec & 0.000073 & 0.00051 \\
2004\ Oct\ 20 & 14:00:00 & 07$^{\rm h}$55$^{\rm m}$52$^{\rm s}$.671889 & +20$^{\circ}$38\arcmin20.56170\arcsec & 0.00001 & 0.0002 \\
2006\ Oct\ 11 & 17:00:00 & 09$^{\rm h}$39$^{\rm m}$54$^{\rm s}$.457138 & +14$^{\circ}$57\arcmin55.39356\arcsec & 0.000023 & 0.00082 \\
2007\ Mar 1  & 07:00:00  & 09$^{\rm h}$31$^{\rm m}$40$^{\rm s}$.709323 & +16$^{\circ}$02\arcmin49.54185\arcsec & 0.000016 & 0.00044 \\
2007\ Jun\ 08 & 00:00:00 & 09$^{\rm h}$31$^{\rm m}$40$^{\rm s}$.531546 & +15$^{\circ}$59\arcmin06.93837\arcsec & 0.000016 & 0.00049 \\
2008\ Jan\ 12 & 10:00:00 & 10$^{\rm h}$41$^{\rm m}$00$^{\rm s}$.869118 & +10$^{\circ}$11\arcmin45.98668\arcsec & 0.000009 & 0.00031 \\
2008\ Jun\ 14 & 00:00:00 & 10$^{\rm h}$22$^{\rm m}$29$^{\rm s}$.258277 & +11$^{\circ}$59\arcmin01.78156\arcsec & 0.000012 & 0.00028 \\
2008\ Aug\ 01 & 22:00:00 & 10$^{\rm h}$40$^{\rm m}$07$^{\rm s}$.840689 & +10$^{\circ}$12\arcmin49.68911\arcsec & 0.000009 & 0.00022 \\
2008\ Nov\ 11 & 17:00:00 & 11$^{\rm h}$24$^{\rm m}$07$^{\rm s}$.553620 & +05$^{\circ}$51\arcmin34.99254\arcsec & 0.000011 & 0.00023 \\
2009\ Feb\ 11 & 14:00:00 & 11$^{\rm h}$27$^{\rm m}$15$^{\rm s}$.292862 & +05$^{\circ}$56\arcmin37.39980\arcsec & 0.000009 & 0.00014 \\
2009\ Apr\ 24 & 06:00:00 & 11$^{\rm h}$09$^{\rm m}$02$^{\rm s}$.825608 & +07$^{\circ}$52\arcmin58.01088\arcsec & 0.000011 & 0.00037 \\
2011\ Feb\ 21 & 13:00:00 & 13$^{\rm h}$04$^{\rm m}$44$^{\rm s}$.278204 & -04$^{\circ}$01\arcmin53.03571\arcsec & 0.000013 & 0.00041 \\
2013\ Mar\ 31 & 11:00:00 & 14$^{\rm h}$33$^{\rm m}$46$^{\rm s}$.748638 & -12$^{\circ}$17\arcmin22.06923\arcsec & 0.000022 & 0.00077 \\
2013\ Jun\ 14 & 07:00:00 & 14$^{\rm h}$14$^{\rm m}$41$^{\rm s}$.930107 & -10$^{\circ}$47\arcmin07.54163\arcsec & 0.000012 & 0.00024 \\
2014\ Jan\ 6 & 18:00:00 & 15$^{\rm h}$15$^{\rm m}$19$^{\rm s}$.189132 & -15$^{\circ}$48\arcmin38.31563\arcsec & 0.000013 & 0.00049 \\

\enddata
\tablecomments{Positions are geocentric at the listed signal reception times.  These  
define the direction vector from the Earth geocenter at signal reception time to Saturn's  
position at signal transmission time (earlier than signal reception by the light travel time 
from Saturn).  Thus, no aberration or relativistic light deflection has been applied.}
\end{deluxetable}

\begin{figure}[!ht]
\includegraphics[angle=0,scale=0.95]{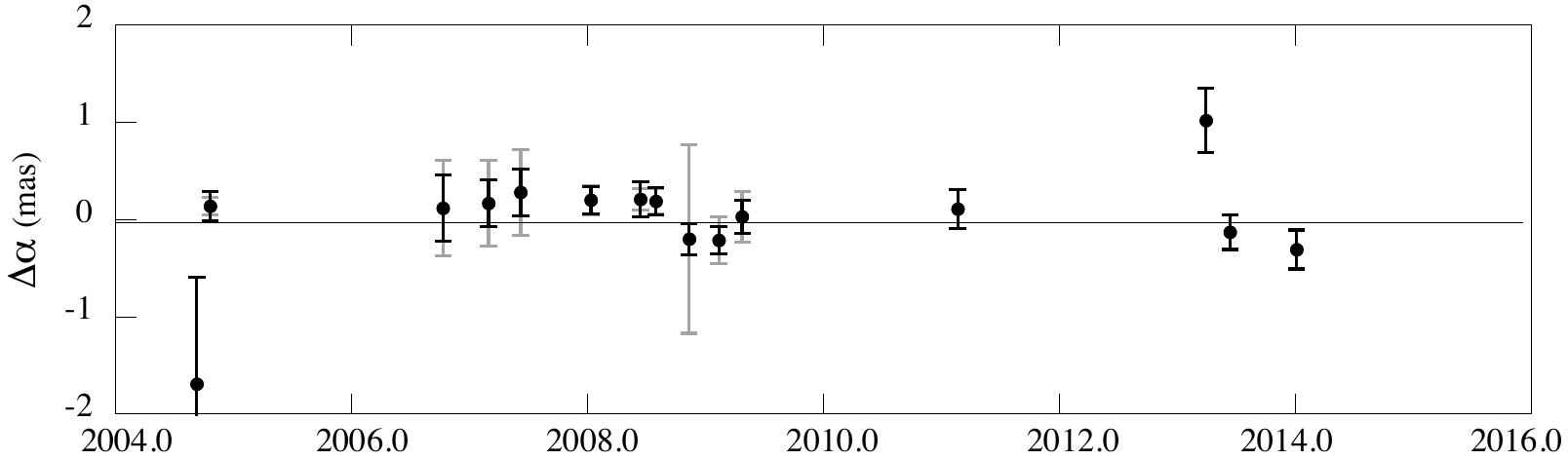} \\
\includegraphics[angle=0,scale=0.95]{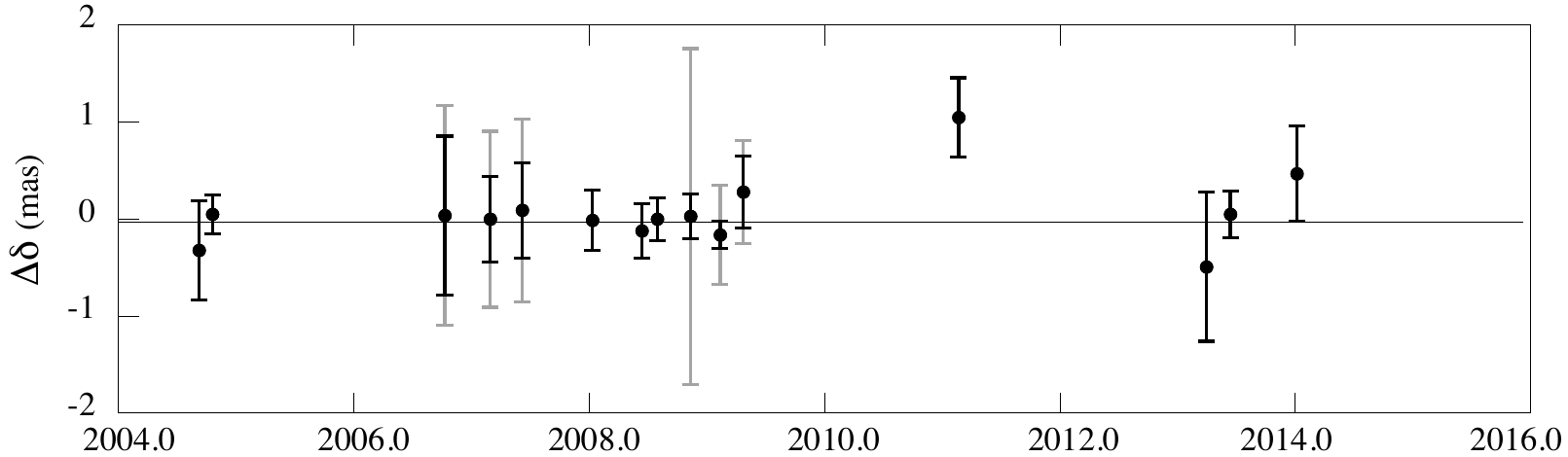}
\caption{Right ascension (top) and declination (bottom) post-fit residuals for VLBA measurements of the position of Saturn, in mas.  
The lighter error bars are from published ICRF and VLBA Calibrator 
Survey (VCS) reference source positions.  These were used previously 
for epochs before 2010.  The darker error bars show the improvement 
when using source positions from the DSN source catalog \citep{Jacobs13}. 
\label{fig2}}
\end{figure}

Figure \ref{fig2} shows the post-fit residuals of our Cassini/VLBA-derived Saturn barycentric positions after fitting to a temporary ephemeris.  A large part of the improvement since the ephemeris fit shown in \citet{Jones11} comes from improved positions for our phase calibration sources.  The rms of the residuals after fitting is 0.5 mas in right ascension using all data, but only 0.2 mas if the two outliers are removed.  One of the outliers is from an epoch prior to Cassini orbit insertion around Saturn when the spacecraft orbit is expected to be less well determined,   The post-fit residual rms in declination is 0.4 mas.  Both rms residual values are consistent with the expected errors.  The larger uncertainty in declination is mainly due to the reduced N-S resolution of the VLBA for low declination sources.  Comparable astrometric results for the European Space Agency's Venus Express spacecraft have been reported by \citet{Duev12} using different antennas and software.

\clearpage
\clearpage

The DE430 planetary ephemeris \citep{Folkner14} is based on fitting the DE421 ephemeris \citep{Folkner09} to our previous Cassini VLBA observations, tracking data from Mars and Venus orbiters (Mars Reconnaissance Orbiter, Mars Express, Mars Odyssey, and Venus Express), and optical observations of the outer planets.  The errors are based on independent estimates of the uncertainties in the VLBA positions including uncertainties in the phase reference source positions, and the orbit determination uncertainties in the position of Cassini with respect to the barycenter of Saturn.  Our more recent Saturn position determinations will help constrain the next generation of JPL planetary ephemerides. 

\begin{figure}[!ht]
\hskip 30pt \includegraphics[angle=0,scale=0.85]{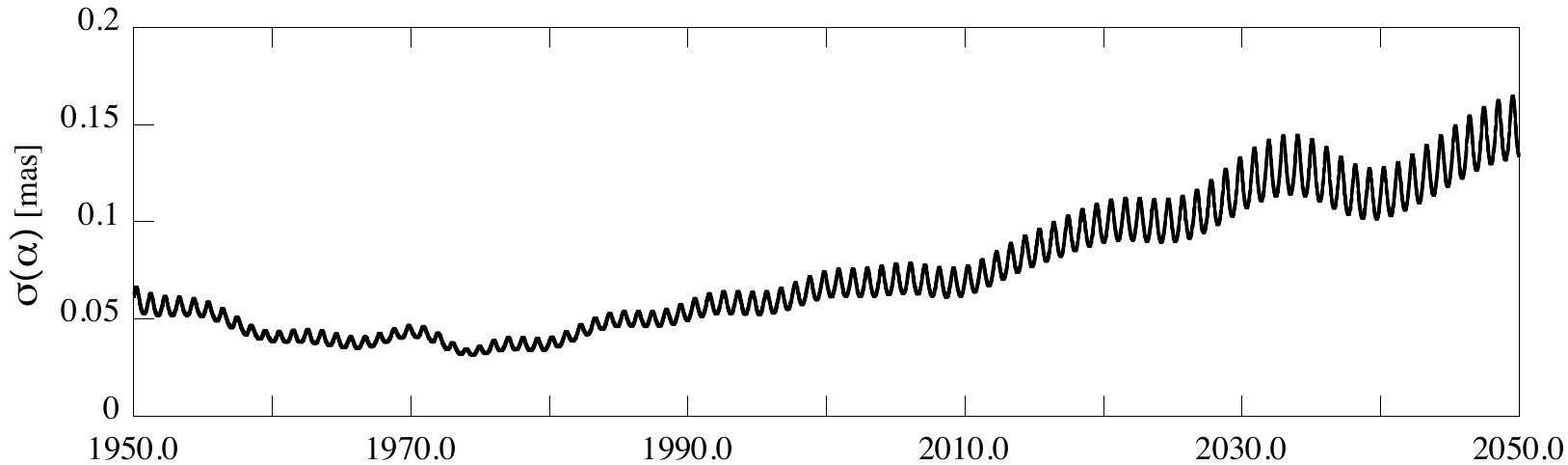}
\caption{Formal estimated uncertainty in the right ascension of the Saturn system barycenter with respect to Earth.
\label{fig3}}
\end{figure}

\begin{figure}[!h]
\hskip 30pt \includegraphics[angle=0,scale=0.85]{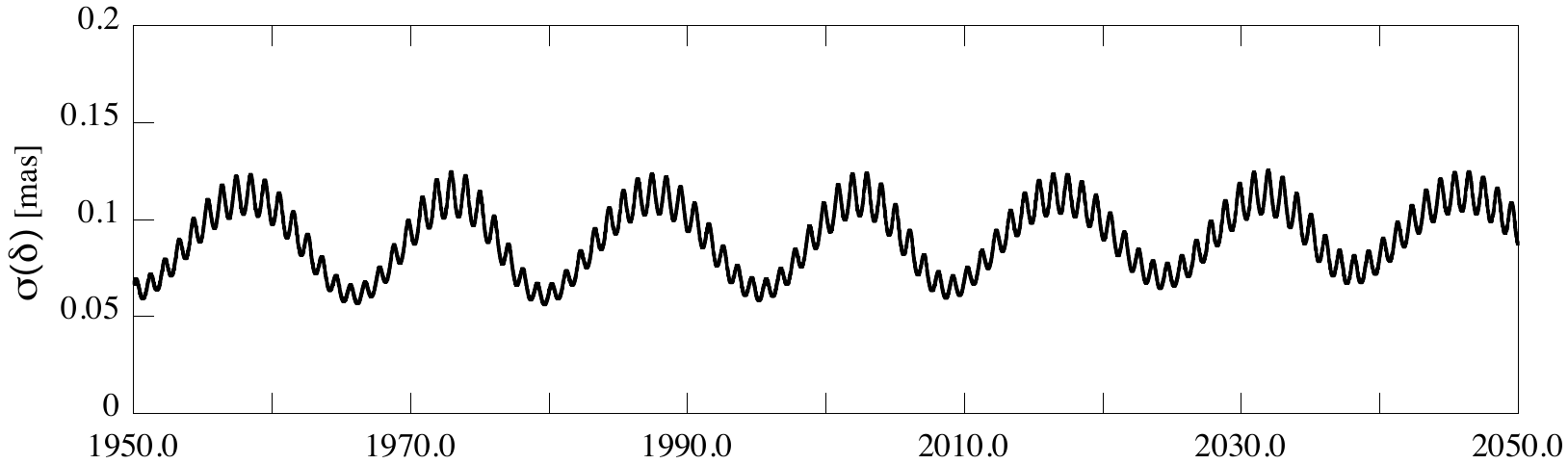}
\caption{Formal estimated uncertainty in the declination of the Saturn system barycenter with respect to Earth.
\label{fig4}}
\end{figure}

\begin{figure}[!hb]
\hskip 30pt \includegraphics[angle=0,scale=0.85]{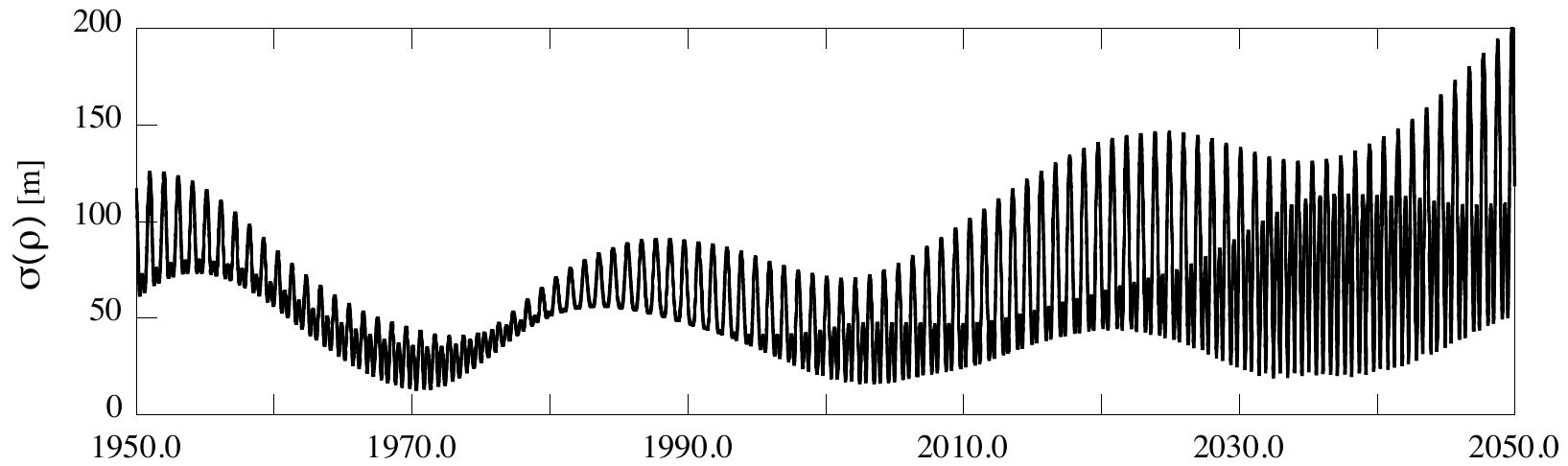}
\caption{Formal estimated uncertainty in the distance from the Saturn system barycenter to Earth. 
\label{fig5}}
\end{figure}

Figures \ref{fig3}, \ref{fig4}, and \ref{fig5} show the formal estimated uncertainty in the right ascension, declination, and distance of the Saturn system barycenter from Earth as a function of time. The uncertainty in declination is determined primarily by the VLBA observations of Cassini described here. The uncertainty in right ascension is determined near the time of the measurements by the VLBA measurements, but increases at later times due to uncertainty in the Saturn semi-major axis, determined primarily by ranging measurements to the Cassini spacecraft \citep{Hees14}. The formal uncertainties shown are typically optimistic since they are based on the assumption that all measurements are uncorrelated. Systematic errors are sometimes common to multiple measurements, such as uncertainty in a quasar location used for multiple VLBA observations, or station delay calibration errors which affect multiple ranging measurements \citep{Konopliv11}. Actual uncertainties in planetary orbits are typically 2-3 times larger than the formal uncertainties.

\clearpage

\clearpage

\section{Conclusions}

The Cassini mission has been extended until 2017.  At that time we will have high accuracy measurements covering one third of Saturn's orbital period.  The error in determining the plane of Saturn's orbit (latitude) rapidly decreases until the time span of observations exceeds 1/4 of the orbital period, while the error in longitude decreases approximately linearly with increasing time span.  Figure \ref{fig4} shows that the average error in declination (latitude) does not grow significantly over a century-long time interval.  The current error for Cassini positions in the ICRF is estimated to be approximately 0.3-0.4 mas (depending on the specific phase reference source used).  Future VLBI observations to maintain and improve the ICRF2 catalog will continue to reduce the position errors of these sources.  

The next mission to a gas giant planet is the Juno mission to Jupiter, launched in August 2011.  The Juno spacecraft will orbit Jupiter for at least one Earth year beginning in July 2016.  This orbiting mission will provide an opportunity to use the same phase referenced astrometry techniques with the VLBA, and thereby improve the ephemeris of Jupiter in a similar manner.   

\acknowledgments

We thank Larry Teitelbaum for past support of this project through the Advanced Tracking and Observational Techniques office of JPL's Interplanetary Network Directorate, and to the VLBA operations staff at NRAO for their continuing excellent support of these observations.  In addition, we gratefully acknowledge support from the NASA Planetary Astronomy Program.  We also thank Peter Antreasian and Fred Pelletier at JPL for providing reconstructed Cassini orbit files used for data correlation at NRAO.  This work made use of the Swinburne University of Technology software correlator, developed as part of the Australian Major National Research Facilities Programme and operated under licence.  Part of this research was carried out at the Jet Propulsion Laboratory, California Institute of Technology, under contract with the National Aeronautics and Space Administration.  

{\it Facilities:} \facility{VLBA}, \facility{Cassini}.

\clearpage

\end{document}